\documentclass[sigconf,letterpaper,nonacm]{acmart}

\setcopyright{none}

\usepackage{xspace}
\usepackage{textcomp}
\usepackage{soul}
\usepackage{marginnote}
\usepackage{hyperref}
\usepackage{amsfonts}
\usepackage[labelformat=simple]{subcaption}
\usepackage{verbatim}
\usepackage[multiple]{footmisc}
\usepackage{comment}
\usepackage{versions}
\usepackage{multirow}
\usepackage{longtable}
\usepackage{makecell}
\usepackage{acro}
\usepackage{adjustbox}
\usepackage{listings}
\usepackage{color}

\def\bitcoin{%
  \leavevmode
  \vtop{\offinterlineskip 
    \setbox0=\hbox{B}%
    \setbox2=\hbox to\wd0{\hfil\hskip-.03em
    \vrule height .3ex width .15ex\hskip .08em
    \vrule height .3ex width .15ex\hfil}
    \vbox{\copy2\box0}\box2}}

\definecolor{dkgreen}{rgb}{0,0.6,0}
\definecolor{gray}{rgb}{0.5,0.5,0.5}
\definecolor{mauve}{rgb}{0.58,0,0.82}

\begin{document}

\title[Dizzy: Large-Scale Crawling and Analysis of Onion Services]{Dizzy: Large-Scale Crawling and Analysis of Onion Services}

\author{Yazan Boshmaf}
\affiliation{%
  \institution{\small{Qatar Computing Research Institute, HBKU}}
  \country{}
}

\author{Isuranga Perera}
\affiliation{%
  \institution{\small{Texas A\&M University}}
  \country{}
}

\author{Udesh Kumarasinghe}
\affiliation{%
  \institution{\small{University of Colombo}}
  \country{}
}

\author{Sajitha Liyanage}
\affiliation{%
  \institution{\small{Qatar Computing Research Institute, HBKU}}
  \country{}
}

\author{Husam Al Jawaheri}
\affiliation{%
  \institution{\small{Qatar Computing Research Institute, HBKU}}
  \country{}
}

\renewcommand{\shortauthors}{Boshmaf et al.}

\begin{abstract}
    With nearly 2.5m users, onion services have become the prominent part of the darkweb. Over the last five years alone, the number of onion domains has increased 20x, reaching more than 700k unique domains in January 2022. As onion services host various types of illicit content, they have become a valuable resource for darkweb research and an integral part of e-crime investigation and threat intelligence. However, this content is largely un-indexed by today's search engines and researchers have to rely on outdated or manually-collected datasets that are limited in scale, scope, or both.
    
    To tackle this problem, we built Dizzy: An open-source crawling and analysis system for onion services. Dizzy implements novel techniques to explore, update, check, and classify onion services at scale, without overwhelming the Tor network. We deployed Dizzy in April 2021 and used it to analyze more than 63.3m crawled onion webpages, focusing on domain operations, web content, cryptocurrency usage, and web graph. Our main findings show that onion services are unreliable due to their high churn rate, have a relatively small number of reachable domains that are often similar and illicit, enjoy a growing underground cryptocurrency economy, and have a graph that is relatively tightly-knit to, but topologically different from, the regular web's graph.
\end{abstract}

\settopmatter{printfolios=true}

\maketitle

\section{Introduction}
\label{sec:intro}

Onion services are private network services that are exposed over the Tor network~\cite{dingledine2004tor}. They allow users to operate and browse web content anonymously without exposing identifying information, such as the network's location or IP address. This anonymity has encouraged onion service operators to host various kinds of contents, ranging from free-speech forums to illicit marketplaces~\cite{spitters2014towards, chen2011dark}.

Onion services were originally developed in 2004 and have recently seen growing numbers in terms of both services and users. As of January 2022, the official Tor Metrics~\cite{tormetrics} count more than 700k onion domains each day, a 20-fold increase from five years ago, collectively serving traffic to 2.5m users at a rate of nearly 8 Gbps. In addition to hosting websites, onion services can be used for instant messaging and file sharing with tools like OnionShare~\cite{onionshare}.

As a breeding ground for many illicit content, onion services have become a valuable resource for security and privacy research. However, finding services with relevant content and collecting high-quality datasets is a very challenging task, limiting the scale or scope of the research. This problem is mainly attributed to the unique properties of onion services, which sets them apart from the regular web. First, they can only be accessed over the Tor network using specialized clients like the Tor Browser. Second, onion domains are hashes of public keys, which make them difficult to remember and link against in websites. Third, the network path between a Tor client and an onion service is typically longer and has a lower bandwidth, increasing latency and reducing the performance of the service. Finally, onion services are ``hidden'' or private by default, which means users have to discover onion domains organically via word of mouth or by surfing linked webpages, rather than relying on search engines.

In this work, we present an open-source system called Dizzy that is designed to crawl and analyze onion services at scale. Dizzy implements novel techniques to (re)crawl onion services and check their status without overloading the guard nodes in the Tor network. In particular, Dizzy uses a carefully organized cluster of Tor clients and webpage rendering engines to serve a highly-distributed cluster of passive crawlers, allowing it to crawl 30 onion webpages per minute per crawler. Moreover, Dizzy uses a distributed filestore, a sharded search engine cluster, and distributed relational/graph databases to enable efficient storage, retrieval, and analysis of crawled webpages, focusing on four aspects of onion services, namely domain operations, web content, cryptocurrency usage, and web graph.

Dizzy groups crawled webpages by onion domains and enriches each domain using a set of cross-validated classifiers, each with an AUC $\ge$ 0.95. This enrichment identifies whether a domain is visually and textually similar to other domains, hosts illicit content, tracks its users, or accepts cryptocurrency payments or donations, in addition to domain category and language detection. Moreover, Dizzy uses specialized hashing techniques to group similar images that are hosted on onion domains, and assigns them to unique source cameras when possible. This enables Dizzy to identify onion domains that host similar images, or images that were captured by the same camera, without having to store likely-illicit image files. In addition, Dizzy creates a directed graph where nodes represent onion or regular web domains and edges represent hyperlinks between them. This graph is used to extract node and graph-level features, in addition to analyzing the connection between onion services and the regular web. For onion domains with cryptocurrency addresses, Dizzy also extracts their corresponding transactions from public blockchains, identifies their wallets using various address clustering heuristics, and finally calculates their money in/out-flows.

We deployed Dizzy in April 2021 and used it to crawl and analyze 63,267,542 webpages from 39,536 onion domains, as of January 31, 2022. We summarize our main findings in what follows:

{\em 1) Domain operations:} Onion services are unreliable. In particular, 47.8\% of the domains were offline on any given day, mainly due to the high churn rate of hosting services. Moreover, 33.1\% of the domains were inaccessible through v3 onion services, and are strictly available through the less secure version of onion services.
    
{\em 2) Web content:} Onion services have a relatively small number of crawler-reachable domains that are often similar and illicit. In particular, 93.0\% of the domains hosted English content and 50.6\% of the domains represented marketplaces. Moreover, 61.7\% of the domains were unsafe and hosted illicit content, especially pornography websites, which almost always contained child sexual abuse material (CSAM). In addition, 75.5\% of the domains were templates and hosted content that is visually and textually similar to at least one other domain. Only 47.0\% of the domains used JavaScript (JS), out of which 6\% have at least one form of JS-based user tracking, especially marketplaces. Finally, 49.3\% of the domains contained at least one Bitcoin address, out of which 41.1\% attribute a total of 56.6k addresses to themselves, typically for making payments or donations. However, 77.7\% of the domains with self-attributed addresses are considered illicit, especially social media websites.

{\em 3) Cryptocurrency usage:} Onion services enjoy a growing underground cryptocurrency economy. In particular, 77.6\% of the attributed Bitcoin addresses were used in 452m blockchain transactions, appearing as inputs, outputs, or both. These 44.0k addresses belonged to 32.6k wallets, which included outliers that had a significantly larger size and money flow, such as exchanges. After filtering out these outliers, nearly 41.7k addresses belonged to 30.4k wallets, which collectively received \$201.5m through 423.4k transactions and sent \$184.0m through 146.2k transactions.
    
{\em 4) Web graph:} Onion services have a topologically different graph structure than the regular web. Considering only onion domains and hyperlinks between them, the corresponding subgraph consisted of 39.5k nodes and 743.2k directed edges. It contained 11.3k strongly connected components (SCC), where the largest component (LSCC) consisted of 40.3\% of the nodes and 44.7\% of the edges. The LSCC had an average clustering coefficient of 0.28, which is nearly half of the regular web's value~\cite{leskovec2009community}. In terms of bow-tie decomposition, the largest component turned out to be an out-component, consisting of 40.3\% of the nodes, whereas in the regular web, the largest component is a core-component, consisting of 27.7\% of the nodes~\cite{broder2000graph}. Finally, 68.0\% of the top-100 domains in terms of their degree centrality were indexers (i.e., link lists), making them the most influential domains in onion services. In terms of connectivity to the regular web, 18.1k onion domains connected to 13.3k regular web domains via 127.4k URLs, out of which 4.1k URLs were malicious. Moreover, 65.9\% of the malicious URLs, which belonged to 864 root-level regular domains, were pointed to by strictly illicit onion services, especially social media websites.

To this end, our main contributions are the design and implementation of Dizzy, its enriched datasets, and the most comprehensive analysis of onion services to date.\footnote{Latest source code and datasets are available upon request. Please contact us via our GitHub organization account found here: \url{https://github.com/cibr-qcri}.} Finally, it is worth mentioning that local authorities are currently using our Dizzy deployment for e-crime investigation and cyber threat intelligence.

\section{Background}
\label{sec:background}

Onion services are TCP-based network services that provide end-to-end security and self-certifying domain names. Given that onion services are only accessible through the Tor network and are mainly used for website hosting, they are often referred to as the darkweb, albeit with a negative connotation.

Each onion service has a unique identifier called an onion domain that is useful only in the Tor network. This identifier, along with other metadata, are packaged into a service descriptor and published to the hidden service directory (HSDir): A distributed hash table where service descriptors can be posted and retrieved. To initiate contact with an onion service, the Tor client uses the onion domain to retrieve the service descriptor from the HSDir, after which the client can establish a path and connect to the onion service.

\begin{figure}
  \centering
  \includegraphics[width=\linewidth]{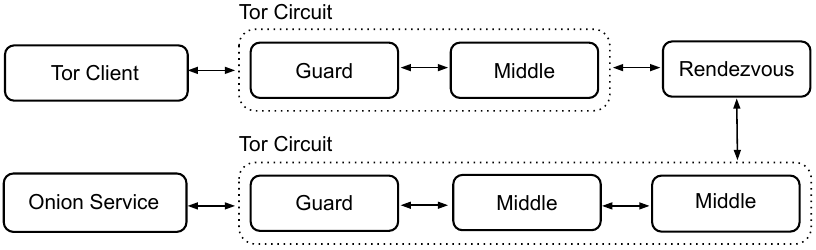}
  \caption{Tor circuits between a client and an onion service.}
  \label{fig:background-tor-circuit}
\end{figure}

Besides end-to-end encryption, onion services provide mutual anonymity (i.e., location hiding), where the Tor client is anonymous to the service and the service is anonymous to the client. By default, a path between a client and an onion service has six Tor relays, two of which are guard nodes representing the entry points to the Tor network. The client builds a circuit to a rendezvous Tor relay, and the onion service builds a circuit to that same relay. As such, no party learns the other's IP address, as shown in Figure~\ref{fig:background-tor-circuit}. The client can change the circuit by resetting the TCP session, for example, but this will only change the middle relays, not the guard node, in order to protect against anonymity-breaking attacks~\cite{dingledine2014one}. The guard nodes typically change every 2–3 months for each client.

To create an onion domain, a Tor client generates an RSA key pair, computes the SHA-1 hash over the RSA public key, truncates it to 80 bits, and encodes the result in a 16-character base32 string. As an onion domain is derived directly from its public key, onion domains are self-certifying and thus provide end-to-end authentication. In other words, if a client knows an onion domain, it automatically knows the corresponding public key. While this property protects the user against impersonation attacks, it unfortunately makes the onion domain challenging to read, write, or remember.

As of February 2018, the Tor Project started deploying v3 onion services: The next-generation onion services whose domains have 56 characters that include a base32 encoding of the onion service's public key, a checksum, and a version number~\cite{holler2021state}. The new onion services also use elliptic curve cryptography, allowing the entire public key to be embedded in the domain, as opposed to only the hash of the public key. Moreover, v3 relies on a new key derivation scheme, where instead of using the onion domain as an identifier for the service descriptor, a new blinded public key is derived from the v3 onion domain and network-specific metadata. As such, no members of the HSDir can read and collect actual onion domains. After three years of deployment, the Tor Browser has disabled v2 onion services support with the release of v0.4.6.x in October 2021. However, it is still possible to access old onion services with other Tor clients, as some of the deployed relays support both versions.

\section{Design}
\label{sec:design}

\begin{figure}
  \centering
  \includegraphics[width=\linewidth]{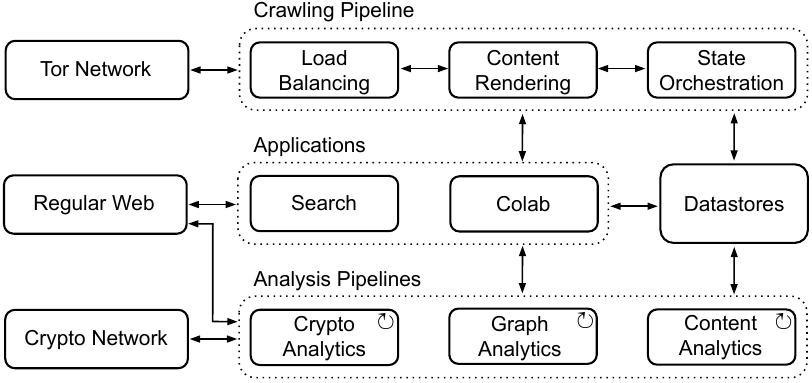}
  \caption{Dizzy's high-level design.}
  \label{fig:design-overview}
\end{figure}

Our key goal is to help researchers and analysts with their darkweb investigations. In particular, we designed Dizzy to overcome a number of challenges related to crawling (\S\ref{sec:design-crawling}), analyzing (\S\ref{sec:design-analysis}), and accessing  (\S\ref{sec:design-apps}) onion services at scale, as shown in Figure~\ref{fig:design-overview}.

\subsection{Crawling}
\label{sec:design-crawling}

Three main challenges are unique to crawling onion services. First, the only way to discover new onion domains is to crawl onion webpages, starting with some known seeds. While one might attempt to collect onion domains by recording directory service queries using modified Tor relays that meet the requirements to join the HSDir~\cite{biryukov2013trawling}, this can only work for v2 onion services, and it is generally considered malicious and impractical~\cite{holler2021state}. Second, as discussed in \S\ref{sec:background}, the Tor client is assigned the same guard node even if it connects to different onion services with their own circuits. As the guard node has a limited bandwidth, it will reject new connection requests once it reaches its full capacity. As such, there is an inherent bottleneck in crawling that introduces a hard limit on performance. Third, onion services host a wide range of illicit contents that are sensitive or illegal, especially CSAM. While common analysis tasks require access to raw data, especially images, it is often prohibited to store such data as part of the crawling process~\cite{kokolaki2020investigating}.

We designed Dizzy with a highly-distributed crawling pipeline that utilizes specialized domain seeding and content hashing techniques to overcome these challenges. In what follows, we discuss each stage of the crawling pipeline in detail.

\subsubsection{State Orchestration}

Dizzy uses three types of crawlers to explore, update, and check the status of onion services. Each crawler type has its own auto-scaling cluster to accommodate increased workloads on-demand, allowing Dizzy to crawl millions of onion webpages in a single day. The crawlers share a job queue containing onion URLs and a hash table that maps visited onion URLs to their metadata, such as rendering parameters. This hash table is used to deduplicate visited URLs from the queue, recrawl visited URLs, and check the status of visited domains. Dizzy relies on a seeding strategy that produces onion domains with diverse contents. In particular, it collects initial seeds by parsing known onion indexers, such as OnionDir,\footnote{\url{http://oniodtu6xudkiblcijrwwkduu2tdle3rav7nlszrjhrxpjtkg4brmgqd.onion}} and the results of search queries from known onion search engines, such as Torch.\footnote{\url{http://xmh57jrknzkhv6y3ls3ubitzfqnkrwxhopf5aygthi7d6rplyvk3noyd.onion}} For the latter source, Dizzy uses search terms that are generated from single words and 2-word combinations from different language dictionaries.

\subsubsection{Content Rendering and Load Balancing}

In Dizzy, HTTP requests are sent through an API provided by an auto-scaling cluster of JS rendering engines. Each renderer then relays requests to a daemon that is part of another auto-scaling cluster of v2/v3-compatible Tor clients, allowing the pipeline to interact with the Tor network using many guard nodes. HTTP responses are sent back by each daemon to the originating renderer for execution, where for each response, it produces a raw HTML file and its rendered version, along with other metadata, such as the response header and hashes of all images found in the rendered webpage, including its screenshot. To utilize images for analysis without having to store them, Dizzy uses difference and perceptual hashing to capture features/scenes of an image~\cite{du2020perceptual} and photo-response non-uniformity (PRNU) noise hashing to fingerprint the source camera used to capture an image, if any~\cite{chen2008determining}. Finally, each rendered HTML file, along with its metadata, are parsed and transformed into a key-value document describing the crawled webpage, and is uniquely identified by its URL. This includes extracting information from the response header, the onion domain, and the HTML markup itself, including URLs, images, JS and Cascaded Style Sheets (CSS) code, either embedded or external, and cryptocurrency addresses. This document is then stored in a sharded search engine cluster, while all remaining files, namely raw/rendered HTML, JS, CSS, and image hash files are stored in a distributed filestore for further analysis. All extracted onion URLs are pushed to the crawling job queue to explore new domains.

\subsection{Analysis}
\label{sec:design-analysis}

Analyzing onion services introduces three main challenges. First, it is common for onion services to host the same website, sometimes with minor changes, under different domain names, typically to improve anonymity and performance. Unlike the regular web, where it is possible to group different domains based on their subject alternative name (SAN) SSL certificate (i.e., a certificate with multiple host names), it is not possible to achieve this grouping at the protocol level in the Tor network due to mutual anonymity, as discussed in~\S\ref{sec:background}. Second, similar to crawling, onion services host a wide range of illicit content that is sensitive or illegal, which makes illicit content detection an essential part of the analysis, typically broken down by domain category. Third, onion services use cryptocurrencies as the default online payment method, mainly due to their privacy features. Unlike centralized methods such as PayPal~\cite{paypal}, where personally identifiable financial transactions are collected and kept private, cryptocurrencies use (pseudo)anonymous identifiers, called addresses, that are shared by onion services to receive payments or donations through public blockchain transactions. In addition, onion services typically include such addresses in their HTML markup as a regular text (i.e., without a unique syntax), making cryptocurrency address attribution to onion services another challenging but essential part of the analysis.

We designed Dizzy with multi-modal analysis pipelines for content, graph, and cryptocurrency analytics of onion services. In what follows, we discuss each pipeline in detail.

\begin{table*}
\centering
\caption{Summary of the extracted features, their type, input date, and target classifiers.}
\small
\begin{tabular}{lclll}\toprule
Extracted feature(s) & Vector? & Type & Input data & Target classifier(s)\\
\midrule
Character 3-gram tokens & \checkmark & Boolean & Cleaned text of rendered homepage & Language \\
\midrule
Uses CSS for page styling & & Boolean & Markup of rendered homepage and CSS files & Illicitness \\
Uses JS for web scripting & & Boolean & Markup of rendered homepage and JS files & Illicitness \\
Number of characters & & Numeric & Cleaned text of rendered homepage &  Illicitness \\
Number of <img> tags & & Numeric & Markup of rendered homepage & Illicitness \\
Number of <button> tags & & Numeric & Markup of rendered homepage & Illicitness \\
Number of <input> tags & & Numeric & Markup of rendered homepage & Illicitness \\
\midrule
Top-10 LDA terms per topic (10 topics) & \checkmark & Boolean & Cleaned text of rendered homepage & Illicitness, Category \\
Number of external URLs & & Numeric & Markup of rendered homepage & Illicitness, Template \\
\midrule
CSS rule sequences & \checkmark & Categorical & Markup of rendered homepage and CSS files & Template \\
Top-10 TF-IDF terms & \checkmark & Categorical & Cleaned text of rendered homepage & Template \\
DOM tree sequences & \checkmark & Categorical & Markup of rendered homepage & Template \\
\midrule
Blacklisted JS functions and API calls & \checkmark & Boolean & Markup of rendered homepage and JS files & Tracking \\
\midrule
Number of cryptocurrency addresses & & Numeric & Markup of rendered webpage & Attribution \\
Address is inside an <a> tag & & Boolean & Markup of rendered webpage & Attribution \\ 
Address is inside a URL & & Boolean & Markup of rendered webpage & Attribution \\ 
Address is inside a <form> tag & & Boolean & Markup of rendered webpage & Attribution \\ 
Address is inside a <footer> tag & & Boolean & Markup of rendered webpage & Attribution \\ 
Address is inside a <table> tag & & Boolean & Markup of rendered webpage & Attribution \\ 
Address is inside <li> or <ul> tags & & Boolean & Markup of rendered webpage & Attribution \\ 
Address is inside an <img> tag & & Boolean & Markup of rendered webpage & Attribution \\ 
Address is located near ``donate'' or ``pay'' & & Boolean & Markup of rendered webpage & Attribution \\ 
Address is located near ``example'' & & Boolean & Markup of rendered webpage & Attribution \\ 
\midrule
Perceptual hash & & Numeric & Images in rendered webpage & Image \\
\midrule
PRNU hash & \checkmark & Numeric & Images in rendered webpage & Camera \\
\midrule
Number of addresses &  & Numeric & Crypto wallet and its transactions & Wallet \\
Number of transactions  &  & Numeric & Crypto wallet and its transactions & Wallet \\
Number of deposit transactions  &  & Numeric & Crypto wallet and its transactions & Wallet \\
Number of withdrawal transactions  &  & Numeric & Crypto wallet and its transactions & Wallet \\
Total value of deposits in USD  &  & Numeric & Crypto wallet and its transactions & Wallet \\
Total value of withdrawals in USD  &  & Numeric & Crypto wallet and its transactions & Wallet \\
Balance in USD &  & Numeric & Crypto wallet and its transactions & Wallet \\
\bottomrule
\end{tabular}
\label{table:desig-analysis-classification-features}
\end{table*}

\subsubsection{Content Analytics}
\label{sec:content_intelligance}

We designed Dizzy with two content analysis sub-pipelines that enable efficient grouping, enrichment, search, and filtering of onion domains by their domain properties and displayed images. This allows Dizzy to lookup services that host content of a given category, for example, or content that includes images which are similar to a given image or captured by the same source camera. We next describe each sub-pipeline in details.

\begin{table}
\centering
\caption{Onion domain categories.}
\small
\begin{tabular}{ll}\toprule
Category & Description \\
\midrule
Social media & A platform for users to share and discuss content\\
Marketplace & An e-commerce website to buy and sell merchandise\\
Pornography & A catalogue of pornographic photos, videos, novels, etc.\\
Indexer &  A search engine or link list for various onion domains\\
Crypto & A service that relies on cryptocurrency transactions\\
Other & A website that does not fit any of the above categories\\
\bottomrule
\end{tabular}

\label{table:design-analysis-classification-categories}
\end{table}

\paragraph{Domain properties}

Dizzy groups documents generated by the crawling pipeline by their onion domains. Instead of processing all of the documents that belong to a domain, Dizzy treats the document representing the homepage of the domain as a representative, and uses it along with its associated files (e.g., rendered HTML, JS, and CSS files) for feature extraction and classification. An exception to this rule is when Dizzy performs cryptocurrency address attribution, where an address can appear on any webpage of a domain, and thus all of the domain's documents and related files may get processed. As summarized in Table~\ref{table:desig-analysis-classification-features}, Dizzy extracts various features for offline training and online evaluation of the following domain property classifiers: 
\begin{enumerate}
\item Language: Identifies the main language of the domain among 50 supported languages using a Naive Bayes (NB) classifier.
\item Illicitness: Identifies whether the domain hosts illegitimate and unsafe content using a Random Forests (RF) classifier.
\item Category: Identifies the main category of the domain among six categories using an RF classifier, as described in Table~\ref{table:design-analysis-classification-categories}.
\item Template: Also referred to as a mirror domain classifier, it identifies whether a domain is visually and textually similar to another using an NB classifier across all pairs of domains.
\item Tracking: Identifies whether a domain tracks its users with JS-based fingerprinting techniques using a Support Vector Machines (SVM) classifier.
\item Attribution: Identifies whether a domain explicitly attributes a cryptocurrency address to itself, as its payment or donation address, using an RF classifier.
\end{enumerate}

The extracted features are stored in a distributed, relational database and are uniquely identified by the corresponding onion domain. These features are updated only if the domain is found to host new content, as indicated by the crawling pipeline during an update run. The trained classifiers and their ground truth datasets are stored in the filestore for subsequent online deployment and retraining. In contrast, classifiers' outputs for each onion domain are stored in the relational database, in addition to updating the corresponding documents in the search engine's index to support custom search filters by domain properties.

\paragraph{Displayed images}

Dizzy processes all image hash files of each domain, both perceptual and PRNU, to identify similar images and source cameras if possible. Before that, however, it filters out all images with size $\le$ 64 pixels, as they typically represent icons, logos, and synthetic images. For the first task, Dizzy uses perceptual hashing as it outputs a similar hash value of an image after it goes through typical transformation and alteration, such as resizing, cropping, blurring, or gamma correction~\cite{du2020perceptual}. In particular, Dizzy uses an image classifier that identifies whether an image is similar to another using agglomerative hierarchical clustering (AHC) across all pairs of images, where hamming distance (HD) is used as a similarity measure~\cite{lux2011content}. As for the second task, Dizzy starts by filtering out all images with size $\le$ 100 $\times$ 100 pixels, as they typically represent previews and thumbnails. Dizzy uses PRNU hashing because each camera creates a highly characteristic pattern caused by differences in material properties and proximity effects during the production process of the camera's image sensor~\cite{chen2008determining}. Specifically, Dizzy uses a camera classifier that identifies whether an image was captured by the same camera that has been used to capture another image using AHC across all pairs of images, where peak to correlation energy (PCE) is used as a similarity measure~\cite{goljan2009large}.

Finally, the outputs of the image and camera classifiers for each onion domain are stored in the relational database, in addition to updating the corresponding documents in the search engine's index to support reverse image search (RIS)~\cite{gaillard2017large}.

\subsubsection{Cryptocurrency Analytics}
\label{sec:crypto-intelligance}

Dizzy runs a cluster of various cryptocurrency daemons to connect to and synchronize with public blockchains, such as Bitcoin and Ethereum. Each daemon represents a full client node with a native RPC API support. Dizzy implements a high-throughput parallel parser that fetches blocks from daemons and then transforms each block, including its embedded transactions and addresses, to a format that is optimized for storage and analysis in the relational and graph databases.

Dizzy clusters the addresses used by each cryptocurrency into wallets using well-known algorithms, such as the multiple-input and deposit address clustering heuristics~\cite{meiklejohn2013fistful,victor2020address}. After that, it filters out outlier wallets that have a significantly larger wallet size and money flow using an Isolation Forest (IF) classifier, even if some of their addresses are self-attributed by onion services. The wallets are then stored in the relational database for further analysis. Dizzy also creates a directed graph where a node represents a wallet and an edge represents one or more transactions whose inputs and outputs contain any of the addresses found in source and destination wallets, respectively. Moreover, each edge has two attributes specifying the number of transactions and the total amount of transferred money. This wallet graph is stored in the graph database to allow efficient money flow-related queries, such as computing the total deposits and withdrawals of a wallet in US dollar (USD).

\begin{table*}
\centering
\caption{Summary of deployed classifiers, their training datasets, and cross-validation performance.}
\small
\begin{tabular}{lllll}\toprule
Classifier & Model & Ground-truth dataset description & AUC & \% (class) \\
\midrule
Category & RF with & 8,881 rendered homepages & $0.99\pm0.01$ & 04.18 (social media) \\
 & One-vs-Rest (OvR) & & & 29.50 (marketplace) \\
 & multiclass strategy & & & 10.81 (pornography) \\
 & & & & 05.81 (indexer) \\
 & & & & 03.83 (crypto) \\
 & & & & 45.87 (other) \\
Language & NB with OvR & 10m Wikipedia abstracts in 50 languages & $0.97\pm0.02$  & 00.02 (each language)\\
Camera & AHC--PCE  with OvR & 2,479 image PRNU hash pairs from 13 cameras &  $0.84\pm0.07$ & 07.69 (each camera) \\
\midrule
Illicitness & RF & 8,881 rendered homepages & 0.97 & 55.91 (illicit) \\
Template & NB & 1,032 pairs of rendered homepages & 0.99 & 13.90 (templated) \\
Tracking & SVM & 1,739 rendered homepages & 0.95 & 35.50 (tracked) \\
Attribution & RF & 2,726 rendered webpages & 0.99 & 57.48 (attributed) \\
Image & AHC--HD & 15,000 image perceptual hash pairs & 0.98 & 43.63 (similar images) \\
Wallet & IF & 1,000 Bitcoin wallets and their transactions & 0.96 & 05.00 (outlier wallets) \\
\bottomrule
\end{tabular}
\label{table:deployment-classification-ground-truth}
\end{table*}

For each cryptocurrency address that has a mapping to an onion domain by the attribution classifier, Dizzy updates the corresponding wallet(s) in the relational database with these attributions as textual labels, in addition to the documents in the search engine's index for fast wallet lookups. In other words, each wallet is labelled by the onion domains which self-attribute any of its addresses.

\subsubsection{Graph Analytics}

Dizzy processes all documents of each domain to construct a directed graph, where a node represents a domain and an edge represents a URL to a domain from another. Moreover, each node has a binary attribute indicating whether it is an onion (type-1) or a regular web (type-2) domain. As such, edges represent URLs to onion or regular web domains from onion domains, where the source node is always a type-1 node. This graph is stored in a distributed graph database that enables fast execution of graph-theoretic algorithms. In particular, Dizzy runs four analytical tasks every time the graph structure changes: Summary statistics, bow-tie decomposition~\cite{broder2000graph}, and centrality measures using type-1 subgraph, and dark-to-regular web linking using the whole graph.

\begin{figure}
    \centering
    \setlength{\fboxsep}{0pt}\fbox{\includegraphics[width=0.95\linewidth]{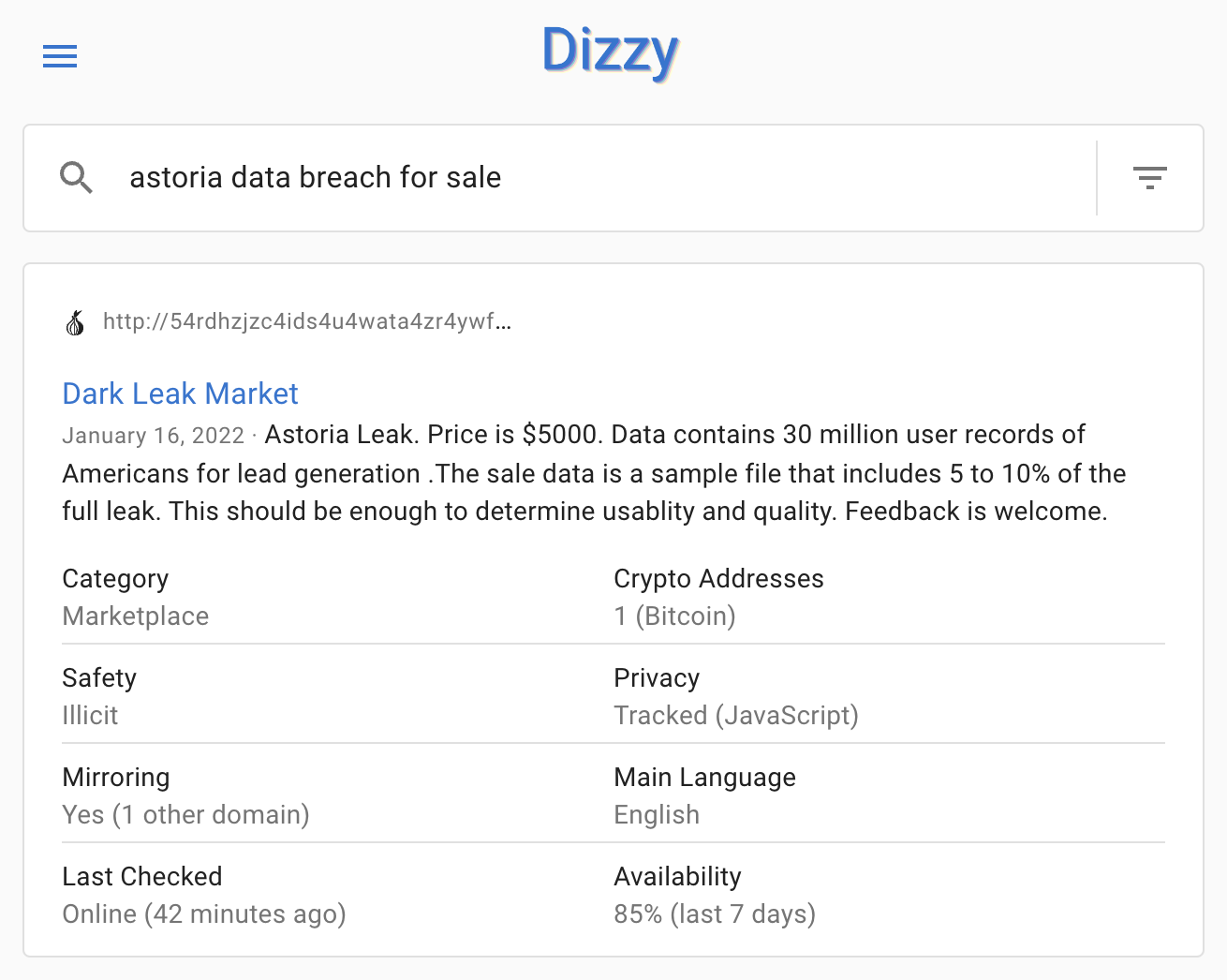}}
    \caption{Dizzy's search web application.}
    \label{fig:analysis-applications-search}
\end{figure}

While the first three analytical tasks are solely graph-theoretic and have existing algorithms, the fourth task involves analyzing onion services interactions with possibly malicious domains on the regular web. To achieve this, Dizzy relies on VirusTotal (VT)~\cite{virustotal} as a malicious domain intelligence feed. In particular, VT provides aggregated URL intelligence by consulting over 70 third-party anti-virus tools and URL/domain reputation services. We refer to each of these tools as a scanner. A primary measure of maliciousness from VT is the number of scanners that mark a URL as malicious. The higher this value is for a given URL, the more likely the URL is malicious. Prior work that relies on VT for malicious URL detection has used a threshold value ranging from one to five~\cite{sharif2018predicting,wang2014whowas,sarabi2018characterizing,nabeel2020following}. As there is no robust way to decide which scanner is more accurate~\cite{zhu2020measuring}, Dizzy flags a URL as malicious if it is flagged by at least one scanner.

The outputs of these tasks are stored in the relational database, where node-specific information, such as its centrality and topological location in the graph, can be used for further analysis.

\subsection{Applications}
\label{sec:design-apps}

Two main challenges in accessing and analyzing crawled datasets are usability and flexibility. While existing technologies like analytical search engines and collaboration laboratories (colabs) represent solutions with desirable trade-offs, they are rarely applied to onion services due to the lack of public, representative datasets. To enable researchers and analysts explore and analyze onion services at will and at scale, Dizzy offers similar technologies for the first time as free-to-use web applications. For example, Figure~\ref{fig:analysis-applications-search} shows a sample search query and result returned by Dizzy's search application.\footnote{\url{https://dizzy.cibr.qcri.org}} 

\section{Deployment}
\label{sec:deployment}

In what follows, we discuss the setup behind Dizzy's real-world deployment and its main ethical considerations.

\subsection{Setup}

We deployed Dizzy in April 2021 on a locally-hosted 35-node Kubernetes cluster with a shared 1Gbps Internet uplink. Each node was configured with a 32-core vCPU and a 64GB VM memory.

We configured Dizzy with 100 Tor client daemons, 100 JS renderers, and 150 crawlers (50 per type), allowing it to crawl 75 onion webpages per second. This also enabled Dizzy to perform daily updates of its index by re-crawling visited URLs, in addition to hourly status checks of each discovered onion domain. For wallet analysis, we configured Dizzy with 10 full-node Bitcoin daemons that enabled parallel parsing. As for the used classifiers, we manually curated ground-truth datasets for offline training and 10-fold stratified cross-validation, as summarized in Table~\ref{table:deployment-classification-ground-truth}.

\subsection{Ethical Considerations}
\label{sec:ethical}

Dizzy's functionality depends on data that is collected and aggregated from public onion services. This data allows Dizzy to offer an easy way to explore, filter out, and search the darkweb for users and services that are relevant to an ongoing investigation. As such, we are faced with privacy concerns related to data collection and analysis, especially knowing that both users and service providers have high expectations of privacy when using Tor. Below we discuss the actions we took to address these concerns in accordance with the guidelines of Tor Research Safety Board~\cite{torsafetyboard}, Tor Ethical Research Guidelines~\cite{torethicalguide}, and our institution's IRB board.

For data collection, Dizzy uses crawlers that target only publicly-available onion domains. The crawlers are polite, passive, and respect robots.txt instructions. This means Dizzy does not collect data from services that require authentication, payment, or email exchange. Dizzy also balances the load across the guard nodes in order to avoid overwhelming the Tor network. In addition, all data collected is secured and stored on our private infrastructure whose access is restricted to authorized researchers.

As for data analysis, Dizzy provides APIs for domain classification and cryptocurrency tracing of onion services, which can result in linking services and their users to particular content types or cryptocurrency transactions/wallets. While such analysis may deanonymize a small fraction of Tor users when combined with regular web data, especially public social media~\cite{al2020deanonymizing}, it does not put users at any additional risk. Instead, it exposes existing risks and corrects common misconceptions, such as Bitcoin being a private or anonymous online payment system.

Finally, we shared Dizzy's deployment plan with our stakeholders to get their early feedback. In response, we engaged with local authorities who were interested in Dizzy and are currently using it for e-crime investigation and cyber threat intelligence.

\section{Results}
\label{sec:results}

As of January 31, 2022, Dizzy has crawled and analyzed 63,267,542 webpages from 39,536 onion domains, where 95.2\% of the domains hosted less than 75 webpages. Moreover, Dizzy has collected over 32m, 2.3m, and 1.5m JS, CSS, and image hash files, respectively, in addition to 57.2k Bitcoin addresses from 514.4k webpages. Finally, Dizzy has parsed Bitcoin's blockchain consisting of 721,240 blocks.

Next, we present the main findings from a 10-month long deployment, focusing on four aspects of onion services, namely domain operations, web content, cryptocurrency usage, and web graph. 

\subsection{Domain Operations}
\label{sec:analysis-service-operations}

\begin{figure*}[!htb]
\centering
\hfill
\minipage{0.33\textwidth}
\centering
  \includegraphics[width=0.9\linewidth]{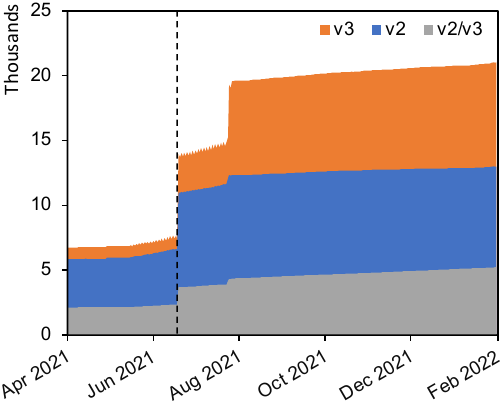}
  \caption{Number of discovered domains}
  \label{fig:analysis-domain-ops-num-domains}
\endminipage
\hfill
\minipage{0.33\textwidth}
\centering
  \includegraphics[width=0.8675\linewidth]{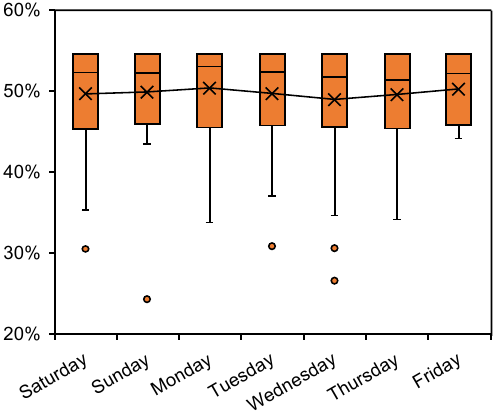}
  \caption{Availability over week days}
  \label{fig:analysis-domain-ops-week-availability}
\endminipage
\hfill
\minipage{0.33\textwidth}
\centering
  \includegraphics[width=0.9125\linewidth]{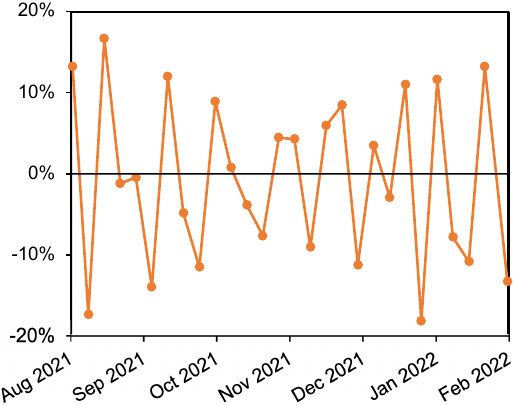}
  \caption{Weekly churn rate}
  \label{fig:analysis-domain-ops-weekly-churn-rate}
\endminipage
\hfill
\end{figure*}

As discussed in~\S\ref{sec:background}, onion services have transitioned to v3 and received various security-related improvements. However, we found that only 53.6\% of the crawled domains operated strictly under v3 by the end of the deployment period, while 33.1\% of the domains still operated strictly under v2, even though the official Tor Browser does not support this version anymore. The remaining 13.3\% of the domains operated under both versions by hosting the same website using v2 and v3 onion services, as identified by Dizzy's template domain classifier (\S\ref{sec:content_intelligance}). Moreover, as shown in Figure~\ref{fig:analysis-domain-ops-num-domains}, we found that most of the discovered v3 domains appeared after June, 2021, which coincided with the official announcement that the Tor Project will no longer support v2 starting with release 0.4.6.x~\cite{torv2deprecation}. On average, Dizzy was able to discover 8.9 new onion domains per day, after excluding the spikes attributed to domains crawled from seeds in the beginning of the deployment and newly discovered indexers in later stages, specifically in June 19 and July 25, 2021.

While the Tor network is reliable, its onion services suffer from low availability. We consider an onion domain available if it has a descriptor in the HSDir and its webserver is up and does not return an error (i.e., HTTP 5xx response code). We found that on average only 52.2\% of the crawled domains were available and mapped to active services on any given day, as shown in Figure~\ref{fig:analysis-domain-ops-week-availability}. However, this low availability of onion domains is mainly attributed to their high churn rate, as shown in Figure~\ref{fig:analysis-domain-ops-weekly-churn-rate}, where a positive value represents the percentage of domains that became unavailable by the end of the week, and vice versa. In terms of week-to-week absolute churn, the largest increase and decrease in number of available domains were 3.5k and 3.8k, respectively.

Similar to the regular web, onion services host their content on popular webservers. We found that 86\% of the crawled domains were served by Nginx, followed by Apache HTTP Server (11\%), Lighttpd (0.2\%), Microsoft-IIS (0.1\%), and a long list of less known and obscure servers. As onion services provide end-to-end encryption, using HTTP over TLS (HTTPs) for onion services is considered redundant if not harmful~\cite{torhttps}. However, we found that 173 of the domains were accessible over HTTPs through 156 TLS certificates, where 17 certificates were reused by one or more domains ($\mu=1$, $\sigma=1.86$, $Q_3=1.25$). In~\S\ref{sec:https-discussion}, we discuss these results in more details and highlight the main use cases of HTTPs for onion services.

\subsection{Web Content}
\label{sec:analysis-web-content}

\begin{figure*}[!htb]
\centering
\hfill
\minipage{0.33\textwidth}
\centering
  \includegraphics[width=0.85\linewidth]{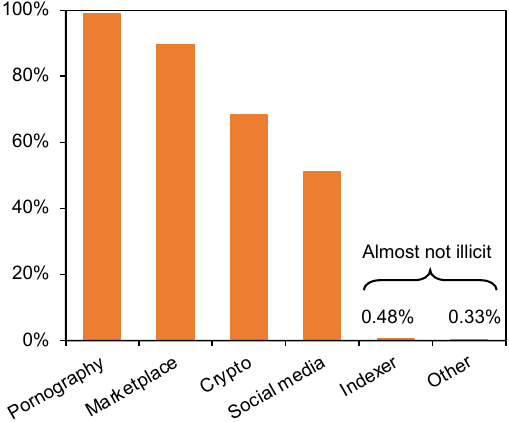}
  \caption{Illicit domains across categories}
  \label{fig:analysis-web-content-category-breakdown}
\endminipage
\hfill
\minipage{0.33\textwidth}
\centering
  \includegraphics[width=0.85\linewidth]{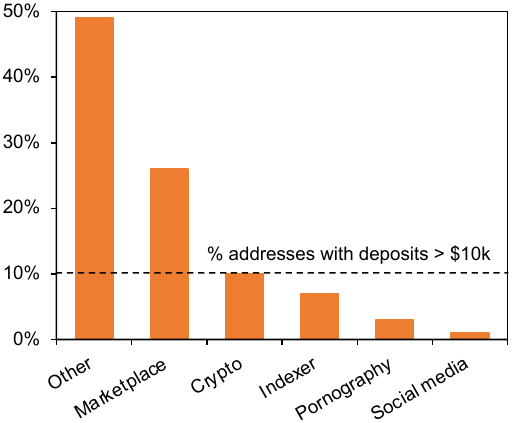}
  \caption{Deposits $>$ \$10k across categories}
  \label{fig:analysis-crypto-deposits}
\endminipage
\hfill
\minipage{0.33\textwidth}
\centering
  \includegraphics[width=0.9825\linewidth]{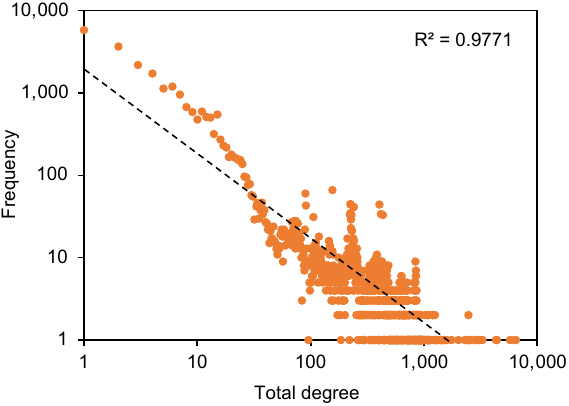}
  \caption{Degree distribution of onion web graph}
  \label{fig:analysis-web-graph-deg-distro}
\endminipage
\hfill
\end{figure*}

As described in~\S\ref{sec:design-analysis}, Dizzy's crawling pipeline processes the homepage of each onion domain at the root path to extract raw/rendered HTML content, in addition to JS, CSS, and image hash files. However, crawling the homepage can result in a server-side redirection with HTTP 302 response code, or a client-side redirection with an HTML <meta http-equiv=``refresh''> tag or a JS window.location object. In this case, Dizzy follows the redirection chain until the last onion URL, after which the corresponding webpage is marked as the homepage if it is hosted under the same source domain. We found that 14\% of the crawled domains resulted in primarily server-side redirections, out of which 99\% did not lead to homepages.

We now present the main findings related to domain properties. First, in terms of language, 93\% of the crawled domains used English as their primary language, followed by Russian (2.1\%), German (1\%), French (0.7\%), and then 40 other identified languages. Second, looking at categories, we found that 50.6\% of the crawled domains were categorized as marketplaces, followed by pornographic websites (9.4\%), cryptocurrency services (7.3\%), indexers (5.4\%), social media like forums (4.1\%), and then ``other'' websites like personal blogs. Third, 61.7\% of the crawled domains were flagged as illicit, albeit indexers and ``other'' onion domains were almost not illicit with only 8 (0.48\%) and 37 (0.33\%) illicit domains, respectively, as shown in Figure~\ref{fig:analysis-web-content-category-breakdown}. On the other hand, almost all pornographic websites were illicit except 47 (1.1\%) domains, which included websites such as PornHub that are popular on the regular web. Fourth, with regards to templates, 75.5\% of the crawled domains were visually and textually similar to one or more domains, forming 2.2k mutually-exclusive clusters of ``templated'' or mirrored domains. The size of these clusters is highly skewed ($\mu$=11.9, $\sigma$=50.9, $Q_3$=4), where the top 10\% contained 77.4\% of the mirrored domains. Fifth, only 47.1\% of the crawled domains used JS, out of which 6.0\% have at least one form of JS user tracking. Moreover, while only 502k (15.7\%) of collected JS files belonged to homepages, 45.4\% of them tracked users, out of which 86.7\% belonged to marketplaces. Sixth and last, 49.3\% of the crawled domains contained at least one Bitcoin address, out of which 41.1\% attributed the addresses to themselves. Moreover, 77.7\% of the attributing domains were flagged as illicit, out of which 53.1\% belonged to social media websites.

In terms of displayed images, we found that 48.7\% of images had a PNG file format, followed by JPG (43.4\%), and GIF (3.2\%). Moreover, 82.4\% of all images had a size $\le$ 64 pixels, which typically represents icons, logos, and synthetic images, and were excluded from further analysis. From the remaining 17.4\%, a total of 267.3k images, we found that only 5.7\% had exchangeable image file format (Exif), which contains metadata tags such as GPS location, camera settings, and image metrics. Moreover, 64.4\% of the images were similar to at least one other image, forming 20.3k mutually exclusive clusters. The size of the clusters is highly skewed ($\mu$=15.3, $\sigma$=61.4, $Q_3$=7), where the largest top 10\% contained 21.3\% of the images. In addition, 71.3\% of the images had a size $\le$ 100 $\times$ 100 pixels, which typically represents previews and thumbnails, and were excluded from further analysis. From the remaining 28.7\%, a total of 76.7k images, we found that they mapped to 13.4k mutually-exclusive source camera clusters ($\mu$=5.6, $\sigma$=33.2, $Q_3$=6), where the top 10\% contained 15.7\% of the images. The distribution of these cameras across onion domains is highly skewed ($\mu$=17.2, $\sigma$=27.4, $Q_3$=4), suggesting that source camera attribution can be a useful technique for filtering domains and narrowing down the investigation.

\subsection{Cryptocurrency Usage}
\label{sec:analysis-crypto-usage}

As described in~\S\ref{sec:design-analysis}, Dizzy processes all documents that belong to each onion domain in order to extract, attribute, and cluster cryptocurrency addresses. We found that 41.1\% of the crawled domains self-attributed 325,653 Bitcoin addresses that are unique per domain. However, the number of unique addresses per domain is highly skewed ($\mu$=40.6, $\sigma$=163.9, $Q_3$=12), where the top 10\% contained 74\% of the addresses. Moreover, only 17.4\% of the attributed addresses were unique across domains, and 77.6\% of them were used at least once as inputs, outputs, or both in 452m transactions. This resulted in a total of 43,971 utilized addresses, which are attributed by domains and used in the Bitcoin blockchian. Finally, 9.1\% of the utilized addresses were flagged as malicious, as they belonged to illicit domains, especially social media websites.

In terms of payments, only 10.2\% of the utilized addresses have received deposits $>$ \$10k, as shown in Figure~\ref{fig:analysis-crypto-deposits}, and only 1.6\% have received $>$ \$100k. These addresses mapped to 32.6k wallets, which included outliers, such as exchanges, bridges, and wallets, with a significantly larger money flow and wallet size (i.e., number of addresses in the wallet). The size of the wallets is highly skewed ($\mu$=5.1k, $\sigma$=200k, $Q_3$=1), where the top 10\% contained 99.9\% of the addresses. Dizzy filters out outlier wallets using its wallet classifier, ending up with 41.7k addresses that belonged to 30.4k wallets. Dizzy also computes the total deposits and withdrawals per wallet in bitcoin (\bitcoin) and USD (\$). For the latter currency, Dizzy uses CoinDesk~\cite{coindeskprice} to convert the input or output value of a transaction from Satoshis to USD based on the exchange rate at the time of the transaction.  Overall, the wallets have collectively received \$201.5m as deposits in 423.4k transactions and sent \$184.0m as withdrawals in 146.2k transactions. Table~\ref{table:discussion-top-wallets-post-filter} shows the top-5 addresses in terms of deposits and withdrawals in USD after wallet filtering, where the volume of a wallet represents the number of transactions that involve any of its addresses.

\begin{table}
\centering
\caption{Top-5 Bitcoin addresses with the largest deposits and withdrawals {\em after} wallet filtering.}
{\small
\begin{tabular}{lrrrrrrrr}\toprule
& \multicolumn{2}{c}{Deposits} & \multicolumn{2}{c}{Withdrawals} & \multicolumn{2}{c}{Wallet}\\
\cmidrule(lr){2-3} \cmidrule(lr){4-5} \cmidrule(lr){6-7}
Address & \# txes & Value (\$) & \# txes & Value (\$)  & Size & Volume \\
\midrule
3C2e... & 113 & 5,069,134 & 33 & 4,956,436 & 54 & 307 \\
36EE... & 718 & 1,065,454 & 2 & 854,522 & 3 & 757 \\
1LSg... & 12 & 725,620 & 8 & 737,795 & 1 & 32 \\
14U2... & 4 & 624,337 & 3 & 624,337 & 11 & 463 \\
1Q6N... & 1 & 457,720 & 1 & 457,720 & 1 & 2 \\
\bottomrule
\end{tabular}
}
\label{table:discussion-top-wallets-post-filter}
\end{table}

\subsection{Web Graph}
\label{sec:analysis-web-graph}

As discussed in in~\S\ref{sec:design-analysis}, Dizzy constructs a directed graph where nodes represent either onion (type-1) or regular web (type-2) domains. Looking at the type-1 subgraph (i.e., the onion services web graph), we found that it consisted of 39.5k nodes and 743.2k edges. It contained 11.3k strongly connected components (SCC), where the largest component (LSCC) consisted of 40.3\% of the nodes and 44.7\% of the edges. The LSCC had an average clustering coefficient of 0.28, which is nearly half of the regular web’s value~\cite{leskovec2009community}. In terms of bow-tie decomposition, the largest component turned out to be an out-component, consisting of 40.3\% of the nodes, whereas in the regular web, the largest component is a core-component, consisting of 27.7\% of the nodes~\cite{broder2000graph}. Finally, 68.0\% of the top-100 domains in terms of their total degree centrality were indexers (i.e., link lists). Figure~\ref{fig:analysis-web-graph-deg-distro} shows the total degree distribution of the type-1 subgraph, where the total degree of a node is the sum of its in and out-degrees. These results suggest that the web graph of onion services is highly influenced by indexers, as they are centrally located in the core-component, which is represented by the LSCC, and have paths to any node in the significantly larger out-component.

In terms of connectivity to the regular web, we found a total of 18.1k onion services (45.8\%) that had 127.4k URLs pointing to 13.3k regular web domains (i.e., unique websites). Most of the onion services represented marketplaces (68.0\%), followed by pornography (12.9\%), social media (5.7\%), crypto (5.2\%), and finally other services (8.2\%). We also found that Blockchain~\cite{blockchaincom}, a popular cryptocurrency blockchain explorer, is the most referred regular website, with about 3.4\% of the URLs pointing to it from onion services. This is followed by the Tor Project (3.1\%), Twitter~\cite{twittercom} (2.9\%), the Bitcoin Project~\cite{bitcoincom} (2.4\%), and then Facebook~\cite{facebookcom} (2.4\%). The reason behind Blockchain's popularity in onion services is that illicit cryptocurrency websites, such as Bitcoin multipliers, list real Bitcoin transactions on their homepage and link them to Blockchain for verification, as discussed in~\S\ref{sec:phising-discussion}. As for malicious regular websites, we found that 4.1k URLs were considered malicious, where at least one VT scanner marked them so, out of which 2.7k URLs (65.9\%) from 864 root-level domains were pointed to by strictly illicit services, which constituted 61.69\% of all linked onion services. Many of these illicit onion services represented social media forums (48\%), followed by marketplaces (26.3\%), and pornography (11.4\%).

\section{Discussion}
\label{sec:discussion}

Our main findings show that onion services tend to have low availability due to their high churn rate, are reachable through a relatively small number of domains that are often similar and illicit, enjoy a growing underground cryptocurrency economy, and have a graph that is relatively tightly-knit to, and topologically different than, the regular web's graph. Next, we discuss some of these findings in details and highlight the main limitations of our work.

\subsection{Identity Verification with TLS}
\label{sec:https-discussion}

As shown in~\S\ref{sec:results}, less than 0.5\% of the crawled domains used HTTPs, and some of these domains have reused the same TLS certificate. The most reused certificate appeared in 19 domains and was flagged as invalid due to hostname mismatch, expect for one domain, which self-signed the certificate.\footnote{\scriptsize Hostname mismatch means the common name to which a certificate is issued does not match the domain name of the service hosting the certificate.}  After manual analysis, we found that this domain belonged to an offshore server hosting service called Impreza Host,\footnote{\scriptsize\url{https://imprezareshna326gqgmbdzwmnad2wnjmeowh45bs2buxarh5qummjad.onion}} which has a similar website on the surface web~\cite{imprezaweb}. The other 18 domains that tried to impersonate Impreza Host were flagged by Dizzy as illicit and categorized as marketplaces. Upon further inspection, we found that they correspond to paid services that offer drugs, hacking, and hitmen through cartel-related affiliates.

In terms of certificate authorities (CAs), 78 of the certificates (50\%) were issued by Let's Encrypt under R3 and Let's Encrypt Authority X3 common names. Other known CAs include DigiCert, Comodo, and DFN-Verein.

Finally, only 21 of the certificates (13.5\%) were valid TLS certificates, and corresponded to non-illicit, popular surface web domains, such as Facebook, BBC, The New York Times, ProRepublica, The Washington Post (SecureDrop), ProtonMail, and Brave Browser, among others. While there are a few use cases where using HTTPs for onion services is desired, such as protecting against phishing and secure cookies leaks of a public website with a complex setup (e.g., serving mixed HTTP and HTTPS content), we found that the main use case is to provide clients, who are still anonymous to services, a secure way to validate the identity of a service. This, for example, is crucial to whistleblowers who ``drop'' classified information to media organizations and need to stay anonymous.

\subsection{Phishing Campaigns with Templates} 
\label{sec:phising-discussion}

\begin{figure}
    \centering
    \fbox{\includegraphics[width=0.95\linewidth]{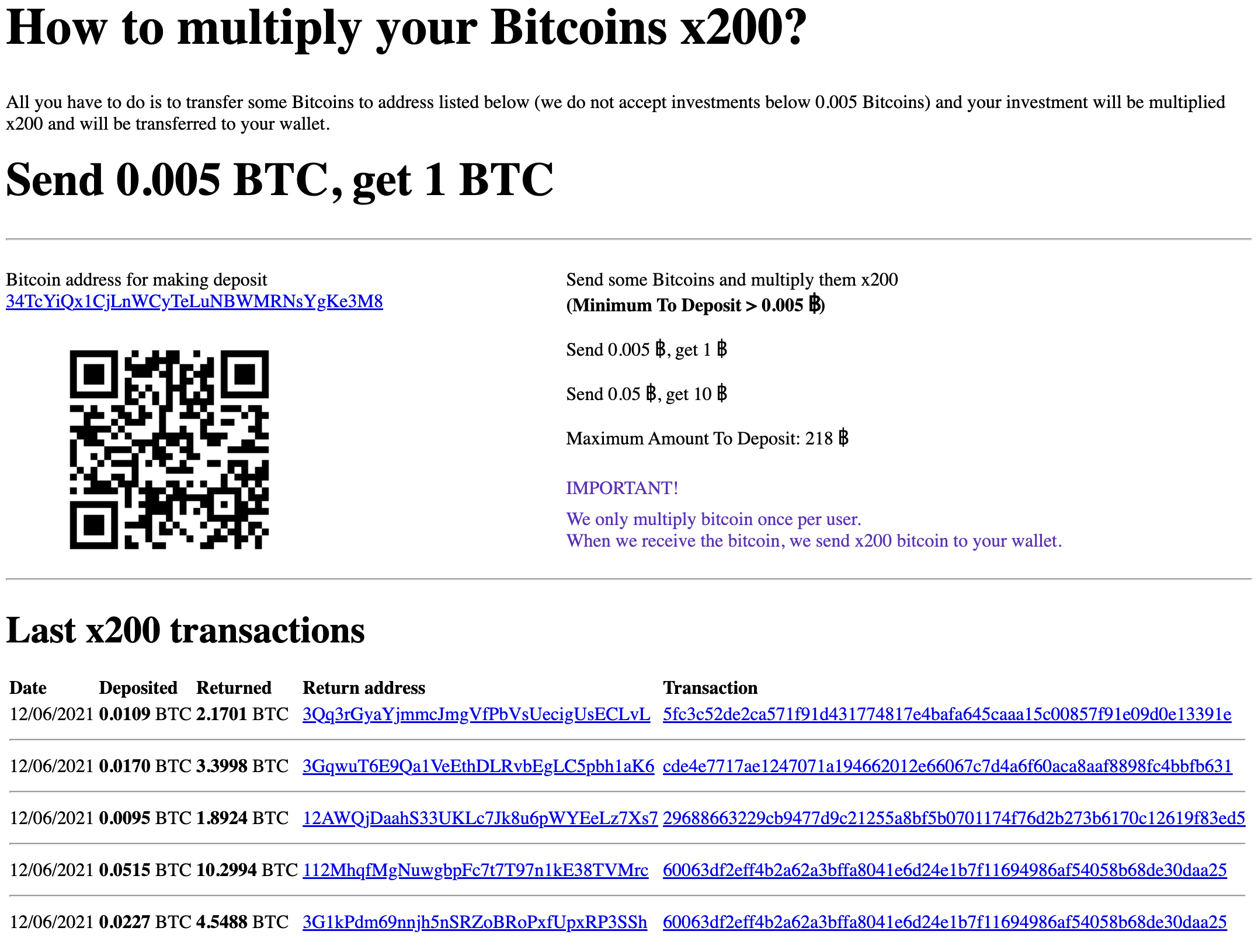}}
    \caption{The mostly used website template on the darkweb.}
    \label{fig:discussion-templates-largest}
\end{figure}

One of the unique features of onion services is the heavy use of templates. As mentioned in~\S\ref{sec:analysis-web-content}, nearly 76\% of the crawled domains were visually and textually similar to at least one other domain, forming clusters of mirrored websites. The mostly used template consisted of 974 unique domains, or 2.5\% of all crawled domains, all of which were flagged by Dizzy as illicit and categorized as cryptocurrency services. After manual inspection, we found that the template represents a fraudulent Bitcoin multiplication website, as shown in Figure~\ref{fig:discussion-templates-largest}. This website uses social engineering to trick users into sending bitcoins to a payment address that is controlled by the website, and in return, the user expects to receive the amount that they have sent multiplied by some factor. Such websites usually falsely claim that they have analyzed the Bitcoin client and found a flaw that allows them to make a transaction in which the recipient would receive more bitcoins than the sender has sent. They even list real Bitcoin transactions that are linked to Blockchain, the cryptocurrency blockchain explorer, as shown in Figure~\ref{fig:discussion-templates-largest}, with ``return'' addresses that received 200x what is claimed to have been sent by users. As such, these website aim to deceive users, and urge them to send bitcoins before that flaw is fixed.

As expected, the domains in the mostly used template hosted nearly identical websites in terms of content, except for some domains which used different payment addresses and showed different recent transactions. This suggests that these websites were automatically generated and hosted by multiple onion services as part of a phishing campaign, in order to increase service availability and reach as many users as possible. We found that 191 of the 974 payment addresses were unique, out of which 5 were utilized Bitcoin addresses. These addresses received a total of \$5,359 in deposits, likely made by multiple victims, through 60 transactions.

\subsection{Reducing Noise with Wallet Filtering}

\begin{table*}
\centering
\caption{Top-5 Bitcoin addresses with the largest deposits and withdrawals {\em before} wallet filtering.}
{\small
\begin{tabular}{lllrrrrrrrr}\toprule
& \multicolumn{2}{c}{Attribution} & \multicolumn{2}{c}{Deposits} & \multicolumn{2}{c}{Withdrawals} & \multicolumn{2}{c}{Wallet}\\
\cmidrule(ll){1-3} \cmidrule(lr){4-5} \cmidrule(lr){6-7} \cmidrule(lr){8-9}
Address & Name & Type & \# txes & Value (\$) & \# txes & Value (\$)  & Size & Volume \\
\midrule
17A1... & Poloniex & Crypto exchange & 459.6k & 950.2b & 436.4k & 950.1b & 1,024.8k & 6.2m \\
1NDy... & Binance & Crypto exchange & 884.5k & 292.7b & 309.4k & 293.0b & 295.0k & 1.2m \\
1G47... & Huobi & Crypto exchange & 1,149.4k & 258.0b & 838.2k & 258.0b& {\em 1} & 1.2m \\
1Kr6... & Bitfinex & Crypto exchange & 342.9k & 101.1b & 276.8k & 100.9b & 813.9k & 2.5m \\
19iq... & RenBTC & Crypto bridge & 59.1k & 5.0b & 59.0k & 4.3b & 35.8k & 0.1m \\
\bottomrule
\end{tabular}
}
\label{table:discussion-top-wallets-pre-filter}
\end{table*}

Onion services may use popular Bitcoin addresses that belong to outlier wallets, such as bridges, exchanges, etc., for various malicious reasons. For example, consider the top-5 Bitcoin addresses with the largest deposits and withdrawals before wallet filtering, as shown in Table~\ref{table:discussion-top-wallets-pre-filter}. The fifth address appears in a post on an onion domain that hosts a hacking forum. In this post, the seller claims that they have compromised the private key of this address, which, as a bridge, is involved in transactions worth billions of dollars. Likewise, the first four addresses in the table, which belong to well-known exchanges, appear in the Bitcoin multiplication phishing campaign discussed in~\S\ref{sec:phising-discussion} as return addresses. In both cases, the goal of using these popular addresses is to deceive users by showing large amounts of bitcoins in real transactions. As such, filtering out outlier wallets is important in order to avoid inflating the results and reaching wrong conclusions.

To show the impact of such outliers, we compared the deposits of the top-5 addresses shown in Table~\ref{table:discussion-top-wallets-post-filter} and Table~\ref{table:discussion-top-wallets-pre-filter}. We found that failing to filter our these addresses will inflate their individual values by a factor of hundreds, which can, for example, drastically over-estimate the size of the darkweb's underground economy.


\subsection{Limitations}
\label{sec:limitations}

While our deployment did not run much into 2022, the 10-month crawling and analysis of onion services resulted in a dataset that is significantly larger and more diverse than previous studies~\cite{platzer2022synopsis}. Moreover, as we wanted to capture changes in the hosted websites before and after v2 was officially unsupported in October 2021, we had to start the deployment a few months before that date and keep it running a few months after it. As a result, we believe that we were able to capture a unique and representative dataset that generalizes recent trends in the darkweb.

Dizzy does not crawl the deep darkweb, that is, the content that sets behind a paywall or requires the user to login, as it uses passive crawlers. While there are legitimate reasons to use active crawlers for deep content crawling of websites, which usually involves creating fake accounts, solving CAPTCHAs, etc., doing so is typically against the EULA and the ethical standards of web crawling~\cite{sun2010ethicality}. Dizzy, however, identifies onion domains that have homepages that require a payment or user login in order to get access to their main content. As such, Dizzy makes it easier for researchers and analysts to identify domains with deep contents, and independently collect data if needed, especially for law enforcement.

The classifiers used by Dizzy, even though they have high AUCs, can result in misclassification and need to be retrained over time. While we did not need to do so in our deployment, we designed Dizzy to handle this situation through active learning with a built-in manual labelling feature at the application level, the search engine web application in particular.

\section{Related Work}
\label{sec:related-work}

In what follows, we discuss related work in two different aspects, namely crawling and analysis, and contrast it to ours. To the best of our knowledge, we are the first to present the most comprehensive analysis of onion services to date, with novel crawling techniques and new analysis results related to domain operations, web content, cryptocurrency usage, and dark/regular web graph topology. 

Biryukov et. al~\cite{biryukov2013trawling} conducted one of the earliest studies on content and popularity analysis of onion services. The authors collected 39k domains by exploiting a flaw in the Tor protocol, the HSDir in particular, to deanonymize and study the popularity of onion services. They analyzed and manually classified the content of 3,050 onion services and found that most popular services are related to botnets. This approach, however, is considered invasive and is no longer feasible with v3 onion services. In contrast, Dizzy uses a passive crawling method that does not abuse any underlying Tor features. It automatically explores, updates, and checks the status of onion services with a load balancing mechanism that does not overwhelm the Tor network, as discussed in~\S\ref{sec:design-crawling}. 

Iskander et al.~\cite{sanchez2017onions} performed a comprehensive analysis of onion services that is partially similar to ours, focusing on graph structure and privacy. The authors crawled 1.5m pages from 7,257 onion domains and analyzed their content, resources, and web graph. They also presented one of the first results on web tracking activities in onion services. Compared to their work, Dizzy crawls and analyzes onion domains at a much larger scale, 65m webpages from 39k domains, through distributed crawling and analysis pipelines. Furthermore, we performed a new set of analyses, including domain operations, multimedia web content, and cryptocurrency usage.

Lee et al.~\cite{lee2019cybercriminal} crawled 27m pages from 37k onion domains and found 5,440 Bitcoin addresses. These addresses were then manually labelled and classified as illicit or not by 10 security researchers, based on the content of the domains in which they were found. Out of all addresses, only 85 were flagged as illicit. The authors then analyzed the money flow from the wallets associated with these addresses, and estimated the volume of darkweb market to be \$180m in terms of deposit. In contrast, Dizzy has collected significantly more Bitcoin addresses, nearly 56.6k addresses from 65m webpages, which were automatically attributed to onion services, classified as illicit or not, and then assigned to filtered wallets, ending up with nearly 44k valid Bitcoin addresses, out of which 4k addresses were illicit. As such, we are able to provide a more accurate estimate of the market's volume at \$201.5m in deposits (or \$184m in withdrawals).

While many studies analyzed the graph structure of onion services~\cite{bernaschi2019spiders,sanchez2017onions,griffith2017graph,aoki2021graph}, they were often limited by a small dataset of onion domains and their URLs, missing a large piece of the onion web graph. For example, Griffith et al.~\cite{griffith2017graph} used a graph that is one-fifth the order of the one we used in this study. Taichi et al.~\cite{aoki2021graph}, on the other hand, presented a more comprehensive analysis using a larger dataset by investigating network changes over time. Similar to these studies, our analysis focuses on summary statistics, bow-tie decomposition, and centrality measures. However, we also analyze the connection between the dark and regular web graphs, and uses the results of these analyses as features in other classifiers (e.g., domain illicitness, templates).

Finally, we note that various studies focused on a single aspect of onion services, such as de-anonymizing darkweb users~\cite{al2020deanonymizing}, detecting illicit domains~\cite{al2017classifying}, identifying forum topics~\cite{tavabi2019characterizing}, and categorizing domains~\cite{ghosh2017automated}. Unlike these studies, our ultimate goal is to provide researchers and analysts with a large-scale, analytical darkweb search engine that facilitates in-depth analysis and investigation of various aspects of onion services. To achieve that, we had to overcome many pitfalls and limitation in darkweb crawling and analysis, as recently highlighted by Platzer and Lux~\cite{platzer2022synopsis} in a survey of 15 papers, by making careful design choices, such as diverse seeding, load balancing, webpage rendering, and image hashing.

\section{Conclusion}

We presented Dizzy: An open-source system for automated crawling and analysis of onion services at scale. The system is designed to help researcher and analysts with their darkweb investigations, focusing on different aspects of onion services. Dizzy is currently deployed in the real-world and is used by local authorities for e-crime investigation and threat intelligence. We used the deployment to perform a comprehensive analysis of onion services, and present novel findings related to onion service availability, domain content similarity, and the connectivity of the dark/regular web graphs.

To this end, we note that we made both the source code and the datasets used in our deployment accessible to interested researchers and analysts from academia or law enforcement agencies, conditional on our institute's approval.

\section*{Acknowledgements}

We would like to thank the folks at the Cybersecurity Initiative for Blockchain Research (CIBR) for their help and feedback.\footnote{\url{https://cibr.qcri.org}} This research was partially supported by Qatar National Research Fund through grant \#QNRF-AICC01-1228-170004.

\setcitestyle{numbers} 

\bibliographystyle{ACM-Reference-Format}
\bibliography{references}

\end{document}